\documentclass[aps,twocolumn,showpacs]{revtex4}
\usepackage{graphicx}   
\usepackage{latexsym}   

\newcommand{\half}{\mbox{${1\over2}$}}               
\newcommand{\etal}{{\em et al.}}                     
\newcommand{\MeV}{{\rm MeV}}                         
\newcommand{\GeV}{{\rm GeV}}                         

\begin{document}

\title{Production of the exotic $\Theta$ baryon 
 in relativistic nuclear collisions}

\author{J\o rgen Randrup}

\affiliation{Nuclear Science Division, 
Lawrence Berkeley National Laboratory, Berkeley, California 94720, U.S.A.}

\date{July 7, 2003}

\begin{abstract}
Recent experimental data appear to confirm the existence of 
the exotic $\Theta^+(uudd\bar{s})$ baryon.
Simple statistical considerations, 
which have generally proven rather successful, 
suggest that this particle should be produced in observable abundance
in relativistic nuclear collisions at RHIC,
where it may be identifed through the $pK^0_s$ invariant mass spectrum.
The observation would not only add new evidence for its existence,
but would also provide an additional means for probing the collision system,
especially with regard to strangeness.
\end{abstract}


\maketitle

\section{Introduction}

Based on an extended chiral soliton model,
Diakonov \etal\ \cite{DiakonovZPA359} made quantitative predictions
for a previously conjectured exotic anti-decuplet of baryons,
the lightest of which would be most easily identifiable.
It can be considered as a $uudd\bar{s}$ pentaquark resonance state
having spin $J=\half$, isospin $I=0$, and strangeness $S=+1$
(hence a charge of $Q=+1$).
In the original paper \cite{DiakonovZPA359},
this particle was denoted $Z^+$
but Diakonov has more recently advocated that
it be named $\Theta$  instead.
We therefore employ this nomenclature.
Its mass was prediced to be about $1530\ \MeV$ 
and its decay width less than $15\ \MeV$.
The dominant decay modes are $\Theta\to K^+n$ and $\Theta\to K^0p$.
Various experiments have recently established
support for the existence of this exotic baryon resonance,
as briefly recalled below.

Nakano \etal\ \cite{Nakano} 
used the Laser-Electron Photon Facility at SPring-8
to irradiate $^{12}$C with photons 
and considered the reaction $\gamma n\to K^- K^+ n$.
They observed a sharp baryon resonance peak at $1540\pm10~\MeV$
in the $K^-$ missing mass spectrum (corrected for Fermi motion).
The Gaussian significance of the peak was reported as $4.6~\sigma$
while the width was estimated to be less then $25~\MeV$.
This evidence strongly suggests the existence of a mole\-cular $KN$ resonance 
with $S=+1$ 
that may be interpreted as the $\Theta$.

Barmin \etal\ \cite{Barmin}
employed a separated $K^+$ beam from the ITEP proton synchroton
to irradiate the bubble chamber DIANA, filled with liquid zenon.
They then considered the charge-exchange reaction $K^+n \to K^0p$
on a bound neutron.
This reaction can be fully measured due to the decay
of the neutral kaon into a $\pi^-\pi^+$ pair inside the chamber.
Their analysis of the $K^0p$ invariant-mass spectrum revealed
a baryon resonance with a mass of $1539\pm2~\MeV$
and a width less than $9~\MeV/c^2$.
The statistical significance was estimated to be $4.4~\sigma$.
This finding is a strong indication for the formation of the $\Theta$.

Most recently,
the CLASS Collaboration has reported results for
photoproduction on deuterium and hydrogen at JLAB \cite{Stepanyan}.
Analyses of the complementary missing mass spectra
yielded a narrow $S=+1$ exotic resonance state in the $nK^+$ system
(with a statistical significance of $5.2\sigma$ and $4.8\sigma$, respectively).
The corresponding mass was found to be approximately $1540~\MeV$.

In view of the mutually consistent experimental evidence 
for the existence of the $\Theta$,
it may be reasonable to consider the implications 
for high-energy nuclear collisions
where baryonic resonances are  abundantly produced \cite{QM01}.
Since the $\Theta$ carries a unique combination of baryon number,
charge, and strangeness it may provide a valuable additional diagnostic tool.

\section{Statistical model}

In order to help ascertain the practical prospects for observing
the $\Theta$ in relativistic nuclear collisions,
we present here a rough estimate of its production yield.
For this purpose we employ statistical considerations
which have been quite successful in accounting for the observed yields
of most hadronic species
(see Refs.\ \cite{Heinz,Becattini,MMRS,Rafelski,MRS}, for example).

In a grand-canonical ensemble of hadron species $i$,
the partition function factorizes into separate contributions for each specie,
$Z=\Pi_i Z_i$, with
\begin{equation}
\ln Z_i(T,V,\{\mu\})=\pm{VTg_i\over2\pi^2}\sum_{n=1}^\infty
{(\pm\lambda_i)^{n}\over n^2} m_i^2 K_2({nm_i\over T})\ ,
\end{equation}
where plus is for bosons and minus is for fermions.
The hadron mass is $m_i$ and $g_i$=$2J_i$+$1$ is the spin degeneracy.
The fugacity associated with the hadron specie $i$ is
\begin{equation}
\lambda_i(T,\{\mu\})={\rm e}^{(B_i\mu_B+Q_i\mu_Q+S_i\mu_S)/T}\ ,
\end{equation}
where $B_i$, $Q_i$, and $S_i$ denote baryon number, electric charge,
and strangeness, respectively.
The ensemble is characterized by the temperature $T$ and
the three chemical potentials $\{\mu\}=\{\mu_B,\mu_Q,\mu_S\}$.
Finally, $V$ is the enclosing volume of the system.

Since the volume is hard to know {\em a priori},
it may be preferable to consider the corresponding spatial {\em densities},
\begin{eqnarray}\label{n1}
n_i(T,\{\mu\})\! &=&\! \pm{g_i T^3\over2\pi^2}\sum_{n=1}^\infty
{(\pm\lambda_i)^n\over n^3}  ({nm_i\over T})^2 K_2({nm_i\over T})~\ \\ 
\label{n2}
&=& {g_i\lambda_i\over2\pi^2}m_i^2 T K_2({m_i\over T})\ \pm\ \dots\\ \label{n3}
&\asymp& g_i \lambda_i \left({m_i T\over2\pi}\right)^{3\over2}
{\rm e}^{-{m_i/T}}\ .
\end{eqnarray}
In the first line, the factors are arranged
such that the terms in the sum are regular in the small-mass limit, $m_i\to0$,
where $K_2$ diverges.
The second line exhibits the leading term,
while the third line shows the simple exponential form
that emerges when the mass is large, $m_i\gg T$.
While we shall use the complete expression (\ref{n1}),
it may be instructive to examine the quality of 
the three levels of approximation exhibited in Eqs.\ (\ref{n1}-\ref{n3}).

Since the sign of the second term in (\ref{n1}) 
depends on the quantum-statistical nature of the particle,
the leading term overestimates fermion densities
while it underestimates boson densities.
As it turns out,
for all the parameter values employed in our yield estimates,
the first term in the expansion, Eq.\ (\ref{n2}),
is an excellent approximation for all baryon species,
being accurate to within a few per mille.
For kaons it is off by a few per cent,
and even the pion densities are understimated by only around $10\%$.
The first term (\ref{n2}) would thus be a quite reasonable approximation
for rough estimates.

By contrast, the large-mass approximation (\ref{n3}),
which generally underestimates the densities,
is never accurate for the situations of interest here.
While underestimating the various baryon densities
by less than $\approx$30\% (only $\approx$20\% for the $\Theta$),
it is off by around half for kaons 
and underestimates the pions by about a factor of five.
Thus it appears that this approximation is less useful.

\section{Estimate of the yield}

We now employ the full expression (\ref{n1})
for estimating the yield ratios that may be expected at RHIC.
We focus on the mid-rapidity region in central Au+Au collisions
at the top beam energy of $100~\GeV/A$,
where the highest freeze-out temperatures have been extracted.
In this environment the net baryon, charge, and strangeness densities
are quite small and,
as one would expect, the extracted chemical potentials are relatively small.
As we shall see, our estimates do not depend much on these parameters.
In order to illustrate this insensitivity,
we employ four different sets of chemical potentials.
They are listed in Table \ref{table-mu} and are motivated below.

For the simpest estimate ($A$),
we assume that all the chemical potentials vanish, 
$\mu_B=0$, $\mu_Q=0$, $\mu_S=0$.
The second set ($B$) uses a finite baryon potential, $\mu_B=30~\MeV$,
corresponding to values for which good fits have been obtained
to the observed hadronic yield ratios \cite{MRS}.
(The sign of $\mu_B$ reflects the overall net baryon number.)
In set $C$ the charge potential has been given a finite value,
$\mu_Q=-4~\MeV$, which leads to good correspondance with 
the results quoted in Ref.\ \cite{MRS}.
(The negative sign of $\mu_Q$ is a consequence of the initial neutron excess
which introduces a corresponding isospin bias.)
Finally, also guided by Ref.\ \cite{MRS},
in set $D$ the strangeness potential is finite as well, $\mu_S=10~\MeV$.
(The fact that a non-zero value of $\mu_S$ is indicated by the data
can be understood as a non-equilibrium feature reflecting
the asymmetry in the elementary strangeness-producing reactions
which causes kaons to be produced more readily than antikaons.)

\begin{table}[h]
\begin{tabular}{|c|cccc|}
\hline
& & & & \\[-2ex]
$\{\mu\}$ set & 
~~~$A$~~~\,\,\ & ~~~$B$~~~\,\,\ & ~~~$C$~~~\,\,\ & ~~~$D$~~~\,\,\  \\[0.4ex]
\hline
& & & & \\[-2ex]
~~$\mu_B$ (MeV)~~    & 0 & 30 & 30 & 30 \\[0.4ex]
~~$\mu_Q$ (MeV)~~    & 0 & ~0 & -4 & -4 \\[0.4ex]
~~$\mu_S$ (MeV)~~    & 0 & ~0 & ~0 & 10 \\[0.4ex]
\hline
\end{tabular}
\caption{\small
The four sets of chemical potentials employed in the calculations,
$\{\mu\}=\{\mu_B,\mu_Q,\mu_S\}$.
}\label{table-mu}
\end{table}

As for the temperature, 
we shall explore a broad range of values, $T=150-200~\MeV$,
in order to bring out the dependence on this important parameter.
(Most fits to the experimental yield ratios
lead to chemical freeze-out temperatures around $175~\MeV$.)
By comparing the results of the different parameter sets,
we may then elucidate the sensitivity to the chemical potentials.

We first calculate the ratio between the $\Theta$ yield
and those of certain familiar hadron species,
namely $\Lambda$, $p$, $K^+$, and $\pi^+$,
which have different combinations of baryon number, charge, and strangeness.
These yield ratios are listed in Table \ref{tableR}
for the four parameter sets $A,B,C,D$:

\begin{table}[tbh]
\begin{tabular}{|cl|cccccc|}
\hline
  ~~$T$ & (MeV)~~ &
~~$150$~\ &~$160$~\ &~$170$~\ &~$180$~\ &~$190$~\ &~$200$~~\ \\[0.4ex]
\hline
& ~$\Theta/\Lambda$  & ~9.0 & 10.7 & 12.5 & 14.3 & 16.1 & 18.0 \\[0.4ex]
$A$ &~$\Theta/p~$    & 3.44 & 4.39 & 5.45 & 6.59 & 7.82 & 9.12 \\[0.4ex]
    &~$\Theta/K^+$   & 0.74 & 1.12 & 1.62 & 2.23 & 2.98 & 3.85 \\[0.4ex]
    &~$\Theta/\pi^+$ & 0.19 & 0.32 & 0.50 & 0.76 & 1.09 & 1.51 \\[0.4ex]
\hline
& ~$\Theta/\Lambda$  & ~9.1 & 10.7 & 12.5 & 14.3 & 16.1 & 18.0 \\[0.4ex]
$B$ &~$\Theta/p~$    & 3.44 & 4.39 & 5.45 & 6.60 & 7.82 & 9.12 \\[0.4ex]
&~$\Theta/K^+$       & 0.90 & 1.35 & 1.93 & 2.64 & 3.49 & 4.47 \\[0.4ex]
    &~$\Theta/\pi^+$ & 0.22 & 0.38 & 0.60 & 0.90 & 1.28 & 1.75 \\[0.4ex]
\hline
& ~$\Theta/\Lambda$  & ~8.5 & 10.2 & 11.9 & 13.7 & 15.5 & 17.3 \\[0.4ex]
$C$ &~$\Theta/p~$~   & 3.35 & 4.28 & 5.32 & 6.45 & 7.66 & 8.94 \\[0.4ex]
    &~$\Theta/K^+$   & 0.90 & 1.35 & 1.93 & 2.64 & 3.49 & 4.47 \\[0.4ex]
    &~$\Theta/\pi^+$ & 0.22 & 0.37 & 0.59 & 0.88 & 1.25 & 1.72 \\[0.4ex]
\hline
& ~$\Theta/\Lambda$  & ~9.1 & 10.8 & 12.6 & 14.4 & 16.3 & 18.1 \\[0.4ex]
$D$ &~$\Theta/p~$    & 3.35 & 4.28 & 5.32 & 6.45 & 7.66 & 8.94 \\[0.4ex]
    &~$\Theta/K^+$   & 0.90 & 1.35 & 1.93 & 2.64 & 3.48 & 4.47  \\[0.4ex]
    &~$\Theta/\pi^+$ & 0.22 & 0.37 & 0.58 & 0.87 & 1.24 & 1.71 \\[0.4ex]
\hline
\end{tabular}
\caption{\small
Hadronic abundance ratios $n_\Theta$/$n_i$ (in per cent), 
with $i=\Lambda, p, K^+, \pi^+$,
for a range of temperatures $T$ (MeV),
as obtained in the grand canoncial treatment (\ref{n1})
with the four sets of chemical potentials $\{\mu\}$ 
listed in Table \ref{table-mu}.
}\label{tableR}
\end{table}

It is evident that the calculated values
show very little variation from one parameter set to the next,
thus rendering the predicted yield ratios
rather insensitive to the precise values of the chemical potentials.
Furthermore, taking the numbers at face value,
the $\Theta$ yield should be about $12-14\%$ of the $\Lambda$ yield.
Consequently,
since good $\Lambda$ data has been obtained,
one would expect the $\Theta$ to be produced at a detectable level.

In order to refine the estimate, 
one may further reduce the sensitivity to the chemical potentials
by considering double abundance ratios 
for which the various attributes are counterbalanced.
The leading fugacity factors then disappear (see Eq.\ (\ref{n2})).
This procedure may be particularly useful with regard to the strangeness
which is not well described by the simple grand canonical model
but requires that an additional suppression factor be introduced \cite{MRS}.
We therefore consider various double abundance ratios for which 
the values of the baryon number, charge, and strangeness balance out.
The corresponding grand canonical results are listed in Table \ref{tableRR}.

\begin{table}[tbh]
\begin{tabular}{|c|c|cccccc|}
\hline
~$\{\mu\}$~ & ~$T$ (MeV)~ & 
~~$150$~\ &~$160$~\ &~$170$~\ &~$180$~\ &~$190$~\ &~$200$~~\ \\[0.4ex]
\hline
$A$ & & 13.7 & 15.6 & 17.5 & 19.4 & 21.3 & 23.2 \\[0.4ex]
$B$ & \underline{$~\Theta/p~$} 
      & 13.7 & 15.6 & 17.5 & 19.4 & 21.3 & 23.2 \\[0.4ex]
$C$ & $K^+/\pi^+$
      & 13.7 & 15.6 & 17.5 & 19.4 & 21.3 & 23.3 \\[0.4ex]
$D$ & & 13.8 & 15.7 & 17.6 & 19.6 & 21.5 & 23.4 \\[0.4ex]
\hline
$A$ &  & 14.1 & 16.0 & 18.0 & 19.9 & 21.8 & 23.7 \\[0.4ex]
$B$ & \underline{$~\Theta/p~$}
       & 14.1 & 16.0 & 18.0 & 19.9 & 21.8 & 23.7 \\[0.4ex]
$C$ &  $K^0/\pi^0$
       & 14.1 & 16.0 & 18.0 & 19.9 & 21.8 & 23.8 \\[0.4ex]
$D$ &  & 14.1 & 16.0 & 18.0 & 19.9 & 21.8 & 23.8 \\[0.4ex]
\hline
$A$ & & ~8.8 & 10.5 & 12.3 & 14.1 & 15.9 & 17.7 \\[0.4ex]
$B$ & \underline{$~\Theta/\Lambda~$} 
      & ~8.8 & 10.5 & 12.3 & 14.1 & 15.9 & 17.7 \\[0.4ex]
$C$ & $K^+/\bar{K}^0$
      & ~8.9 & 10.5 & 12.3 & 14.1 & 15.9 & 17.8 \\[0.4ex]
$D$ & & ~8.8 & 10.5 & 12.3 & 14.1 & 15.9 & 17.7 \\[0.4ex]
\hline
\end{tabular}
\caption{\small
Double hadronic abundance ratios (in per cent) 
for a range of temperatures $T$,
as obtained in the grand canoncial treatment
with the four sets of chemical potentials $\{\mu\}$ 
listed in Table \ref{table-mu}.
}\label{tableRR}
\end{table}

For these double yield ratios,
the dependence on the chemical potentials is practically negligible
and the temperature dependence is considerably smaller
than for the single ratios shown in Table \ref{tableR},
thus making the ``predictions'' more robust.

The first ratio shown,
$(n_\Theta/n_p)/(n_{K^+}/n_{\pi^+})$,
is perhaps the most useful since it involves, in addition to the $\Theta$,
positively charged hadrons that can be directly detected.
Since the experimental yields of those are well known,
the tabulated numbers can be used to make absolute predictions for 
the $\Theta$ yield in terms of the observed yields of $\pi^+$, $p$, and $K^+$.
Inserting the multiplicities reported by BRAHMS \cite{Bearden}, we then find, 
at mid rapidity for central Au+Au collisions at the top RHIC energy,
\begin{equation}
({dn_\Theta\over d\eta})_0 \approx 
{n_\Theta/n_p \over n_{K^+}/n_{\pi^+}}\
{(dn_{K^+}/d\eta)_0 \over (dn_{\pi^+}/d\eta)_0}\ 
({dn_p\over d\eta})_0 \approx 0.9\ ,
\end{equation}
{\em i.e.}\ about one $\Theta$ per unit rapidity in each central event.

The second ratio, $(n_\Theta/n_p)/(n_{K^0}/n_{\pi^0})$,
is included primarily for illustrative purposes.
It has pratically the same value as the first ratio,
as should be expected from the very small isospin dependence of the yields.
But it is less readily observable
(while the neutral kaons can be identified
through their decay via $K_s^0$ to charged pion pairs,
the neutral pions must be measured 
by means of their decay into photon pairs).

The third reaction shown, $(n_\Theta/n_p)/(n_{K^+}/n_{\bar{K}^0})$,
is actually easier to measure since $p$ and $K^+$ can be detected directly,
thus leaving only the task of identifying $n_{\bar{K}^0}$
through the $K_s^0\to\pi^-\pi^+$ decay.
(This identification is anyway required for the observation of the $\Theta$,
see below.)
Alternatively, because of the very weak isospin dependence,
one might fairly safely approximate $n_{\bar{K}^0}$ by $n_{K^-}$.

While the focus of the present brief note is on the top RHIC energy,
where the yield is expected to be largest,
the results given in Table \ref{tableRR} may be useful
for other scenarios as well
since the double yield ratios depend primarily on the freeze-out temperature,
with very little sensitivity to the various chemical potentials.

\section{Discussion}

As the above estimates suggest, the $\Theta$ should be produced 
with non-neglible rates in high-energy nuclear collisions.
It may thus be worthwhile to extract its yield from the data.
Obviously,
the experimental observation of this exotic baryon in nuclear collisions
would provide entirely independent evidence for its existence.
Moreover, if experimentally detectable,
the $\Theta$ would provide an additional means of probing the system,
due to its unique combination of attributes.
Especially useful may be the fact that it is a $S=+1$ baryon,
which could help improve our understanding of strangeness production.

One possible method of analysis for identifying the $\Theta$
might be to consider the invariant-mass spectrum of the $K_s^0p$ system, 
as in Ref.\ \cite{Barmin}.
In addition to the readily available kinematic information on the proton,
this approach requires the $K_s^0$ mesons to be identified
through their $\pi^-\pi^+$ decay vertex.
This should be possible as well with the existing instrumentation
(using the STAR TPC, for example).
Presumably, since the intrinsic $\Theta$ decay width is so small,
the observed width would primarily reflect the detector resolution
(the construction of the $p\pi^-\pi^+$ invariant-mass spectrum
requires complete kinematical information for all three final hadrons).
To the degree that the associated instrumental tolerances are well understood,
it should then be possible to predict the observable width,
thus facilitating the identification of the signal.

The $\Theta$ may also be of interest on the theoretical side
since microscopic transport models
usually attempt to include all known hadronic resonances.
In these treatments,
a primordial hadronic population generated by model-specific mechanisms 
undergoes further evolution involving both decays and reactions.
As a result,
both the thermodynamic properties of the hadronic resonance gas
(such as pressure and entropy) as well as the final yield of
the detectable hadron species 
are affected by the selection of included resonances.

Thus, insofar as the existence of the $\Theta$ baryon resonance
can be considered established,
it would seem natural that it be included in those treatments.
While its expected relatively small abundance
will likely render it ineffective with regard to modifying the overall
thermodynamic conditions,
it may play a role for the abundance of other relatively rare species.
In particular, it might be of interest to investigate
how strangeness-related quantities might be affected
by the presence of the $\Theta$.

Just as this note was being finalized,
it was hypothesized \cite{Capstick}
that the $\Theta$ may in fact not be an isoscalar but rather is an isotensor
having five members.
Three of them, $\Theta^{++}$, $\Theta^{+}$, and $\Theta^{0}$, 
would have isospin violating strong decays,
while the remaining members, $\Theta^{+++}$ and $\Theta^{-}$
would have weak decays and so be long-lived.
This intriguing possibility gives the search for $\Theta$ production 
in nuclear collisions even larger urgency,
since it could quite possibly settle this issue.
Indeed, assuming that $\Theta^{++}$ is produced at about the same rate as
$\Theta^{+}$, as the statistical model would suggest,
it should be readily identifiable as a peak 
in the $pK^+$ invariant mass spectrum
(a task simpler than identifying the $\Theta^+$).

\section{Summary}

Motivated by the recent apparent experimental confirmation
of the predicted exotic $S=+1$ baryon $\Theta(1540)$,
we have briefly discussed its relevance to high-energy nuclear collisions.
Obviously, if this particle indeed exists,
it would enrich the hadronic dynamics and ought to be included in models 
treating the hadronic resonance stage of the collisions.

In order to provide a rough estimate of the expected yield,
we have invoked the framework of statistical equilibrium,
which has been found to account very well 
for the observed hadronic abundancies over a wide range.
These simple estimates suggest that the $\Theta$
may be produced at a level of approximately $6\%$
relative to the proton yield
and about $14\%$ of the $\Lambda$ yield,
rather independently of the prevailing chemical potentials
(as long as their values are reasonably consistent 
with the observed yield ratios).

Further insensitivity to the somewhat poorly understood 
strangeness suppression effect can be obtained by taking 
double yield ratios designed to ensure counterbalancing
of the various conserved attributes (baryon number, charge, and strangeness).
By this method, using the observed $\pi^+$, $K^+$, and $p$ multiplicities,
we estimate that about one $\Theta$ per unit rapidity
will be produced at mid rapidty in central Au+Au collisions at $100~\GeV/A$.

Since the expected $\Theta$ yield is thus not negligible,
we encourage an analysis of the data
for the purpose of establishing $\Theta$ production in heavy-ion collisions.
This task could be carried out by examining the invariant mass spectrum
for the $pK_s^0$ system,
with the $K_s^0$ being measured by means of its $\pi^-\pi^+$ decay vertex.
In view of the possible existence of an entire $\Theta$ multiplet
\cite{Capstick},
it might be worthwhile to examine also the $pK^+$ invariant-mass spectrum
at the same time.

The observation of such a resonance in high-energy nuclear collisions
would provide independent supportive evidence 
for the existence of this exotic particle.
Once its existence has been established,
the $\Theta$ may offer an interesting additional means
for probing the properties of the collision system
(especially those related to strangeness)
due to its unique combination of baryon number, charge, and strangeness.

\section*{Acknowledgements}
This work was supported by the Director, Office of Energy Research,
Office of High Energy and Nuclear Physics,
Nuclear Physics Division of the U.S. Department of Energy
under Contract No.\ DE-AC03-76SF00098.
It was carried out at the National Institute of Nuclear Theory in Seattle.
I wish to acknowledge helpful discussions with 
H.\ Caines, D.\ Diakonov, D.\ Magestro, G.\ Kunde, 
S.\ Stepanyan, R.\ Vogt, and N.\ Xu.



                        \end{document}